\documentclass[12pt,a4paper]{article}
\usepackage{amsmath,amscd,amssymb,amsfonts,amsthm}

\DeclareMathOperator{\Ber}{Ber}
\DeclareMathOperator{\Tr}{Tr}
\def\a{\alpha}

\def\n {{\bf n}}
\def \vol {{\rm vol\,}}
\def\p {{\partial}}

\def\a {\alpha}
\def \E {{\mathbb E}}
\def\S{{{\mathcal S}}}
\def\v{{\bf v}}

\newtheorem{proposition}{Proposition}

\title{Tube formula, Berezinians, and Dwork formula}

\newcommand{\myaddress}{\begin{center}
\small School of Mathematics\\ \small The University of
Manchester\\  \small PO Box 88, Manchester M60 1QD, UK\\
\small {\tt khudian@manchester.ac.uk}
\end{center}}

\author{Hovhannes M.\,Khudaverdian}

\date{\myaddress}

\begin{document}

\maketitle

\maketitle


\begin{abstract}
{\small We consider an example of tubes of hypersurfaces in
Euclidean space and generalise the tube formula to supercase. By
this we assign to a point of the hypersurface in superspace a rational
characteristic function. Does this rational function appear when we
calculate the $\zeta$-function of an arithmetic variety?}
\end{abstract}

\label{first}
   I would like to make a remark on  relations between the tube formula
   and Dwork formula for $\zeta$-function for arithmetic varietes.
  For several years I have been thinking about this relation
   and have discussed it with many  colleagues.
   In particular, I spoke about it in Bia{\l}owie\.{z}a
   last summer\footnote{This note is based on my talk on the XXV-th workshop
   on Geometric Methods in Physics in Bia{\l}owie\.{z}a (July 2006).}.
  Recently a very interesting  paper \cite{candelas} appeared in the web,
   which touches on a related circle of ideas.

\section {Tubes of hypersurfaces}

   Recall some simple facts concerning tubes of hypersurfaces in Euclidean space.

   Let $M$ be a surface in Euclidean space $\mathbb E^{n+1}$.  By a tube  we shall
   understand
    the set of points in $\E^{n+1}$ that are at distance $h$
   from $M$, $h\geq 0$.  If $M$ is an orientable hypersurface (surface of codimension $1$), then
   a  direction of normal vector can be chosen. This defines
   sign of the distance between a point and
   the surface.  In such a case the tube of radius $h$ is the disconnected union of two
    {\it half-tubes} $M_h$ and $M_{-h}$.
   We consider here only oriented hypersurfaces and later denote by $M_h$ a half-tube
   for any $h\in \mathbb R$.
  The $n$-dimensional volumes of tubes and half-tubes
  are polynomials in $h$ if $h$ is small enough.
 These formulae can be traced to Steiner (1840), who derived them
 for a polygon and a polyhedron. In 1939 Weyl gave general formulae for polynomials expressing
 volumes of tubes and half-tubes. The coefficients of these polynomials
 are integrals of expressions which are formed from the second quadratic form at $n$-dimensional
  surface.
 For tubes (not half-tubes) these coefficients {\it do not change} under isometries of the surface;
 they are expressed via internal curvature tensor.
 (An excellent exposition on tube formula containing full references is given  in \cite {grey}).

   Consider first a toy example. Let $M$ be the boundary of a convex polygon.
Then it is evident that $\vol M_h=\vol M+2\pi h$  and $\vol (M_h\cup
M_{-h})=2\vol M$. Henceforth the volume of  a $k$-dimensional
surface $M$ is denoted  $\vol M$. (If $M$ is $1$-dimensional,
  then $\vol M$ is length, if $M$ is $2$-dimensional,
  then $\vol M$ is area.)

Now let $M$ be a closed orientable hypersurface in $\E^{n+1}$ and
$\n$ be normal unit vector field of $ M$.
  Consider new coordinates $(u^1,\dots,u^n,t)$ in a tubular neighborhood of $M$
  defined by the relations
  $x^a(u,t)=x^a(u)-tn^a(x(u))$, where $x^a=x^a(u)$
  is a local parameterisation of $M$.
  Straightforward calculations show  that the Jacobian of transformation from Cartesian
  coordinates $(x^1,\dots,x^{n+1})$  to
   these new coordinates
     $\,\,J=\det\left({\p\left(x^1,\dots,x^{n+1}\right)\over\p\left(u^1,\dots,u^{n},t\right)}\right)=$
                             \begin{equation}\label{jacobianoftransformation}
                      =
                   \det \left(
                    {\p x^a(u)\over\p u^i}-
                    t{\p n^a(x(u))\over \p u^i}, n^a(x(u))
                    \right)=\sqrt {\det g_{ij}(u)}\det (1+tS(u))\,.
                             \end{equation}
Here $g_{ij}={\p x^a\over \p u^i}{\p x^a\over \p u^j}$ is induced
Riemannian metric on $M$ (the first quadratic form). It defines volume
form $d\sigma_u=\sqrt {\det g_{ij}(u)}d^n u$ on the surface $M$ in
the parameterisation $x^a(u)$.
 The linear operator $S$ is variant of second quadratic form.
 It is Weingarten (shape) operator
 defined by the relation $S{\bf v}=-\p_{\bf v}\n$
   for an arbitrary tangent vector $\bf v$: $S^j_i {\p x^a\over \p u^j}=-{\p n^a\over \p  u^i}$ and
   $S^j_i=g^{jk}n^a{\p^2 x^a\over \p u^k \p u^i}$.
(Henceforth we will not distinguish between upper and lower
indices in Euclidean space and we suppose summation over repeated
indices.)

  Using the Jacobian \eqref{jacobianoftransformation} one can easy calculate the volume of the
  half-tube $M_h$
    for small $h$. Let $\rho(t)$ ($t\in \mathbb R$)
     be an arbitrary (smooth) function such that it vanishes outside of sufficiently large
    neighborhood of zero.  Consider the integral
    $\int \rho\left(t(x)\right)d^{n+1}x$,
     $t(x^a)$ being the distance between the point $(x^1,\dots,x^n)$ and surface $M$.
     On one hand this integral is equal to $\int \rho(t)\vol M_t dt$. On the other hand,
     by using formula \eqref{jacobianoftransformation} we arrive at
         \begin{equation}\label{areawithweight1}
         \int \rho(t(x))d^{n+1} x =\int  dt\rho(t)\left(\int_{M}\det (1+t S(u))d\sigma_u\right)\,.
      \end{equation}
In particular it follows that if $M$ is closed and $h$ is small enough then
 the volume of half-tube
is equal to
       \begin{equation}\label{areaofpositivepart}
    \vol M_h=\int_{M}\det (1+h S(u))d\sigma_u=\sum_{k=0}^n c_k h^k\,.
   \end{equation}
  Coefficients $c_k$ are as follows:
 $c_0=\int_{ M} d\sigma=\vol ( M)$, $c_1=\int_{ M} \Tr Sd\sigma$, the integral of mean curvature over surface,
etc.  The last coefficient $c_n=\int_{ M} \det Sd\sigma$ is equal to
 the volume of unit $n$-dimensional sphere
multiplied by the  degree of Gaussian map $M{\buildrel \n (u)\over
\longrightarrow S^{n}}$. (If $n$ is even then the degree  is equal up to a
factor to Euler characteristic $\chi(M)$). E.g. if $M$
is two-dimensional closed (oriented) hypersurface in $\E^3$, then
$\det(1+tS)=1+t \Tr +t^2\det  S=1+t H+t^2 K$, where $H=k_1+k_2$ is
mean curvature, $K=k_1k_2$ is Gaussian curvature ($k_{1,2}$ are principal curvatures). The
volume (area) of
half-tube $M_h$ is equal to $\vol M_h=\vol M+h\int_M H
d\sigma+h^2\int_M K d\sigma=\vol M+h\int_M H d\sigma+2\pi \chi
(M)h^2$. Respectively, the volume of the tube $M_h\cup M_{-h}$ is
equal to $2\vol M+4\pi \chi (M)h^2$.

  Summarizing we can say the following:
    To an arbitrary hypersurface $M$ in $\E^{n+1}$ one can assign
  {\it a local characteristic polynomial}
\begin{equation}\label{charpollocal}
    P_M(t,x(u))=\det (1+tS(u))
\end{equation}
and its integral over the surface,
  {\it the characteristic polynomial}
  \begin{equation}\label{charpol}
    P_M(t)=\int_M  P_M(t,u)d\sigma_u=\int_M  \det (1+tS(u)) d\sigma_u\,.
\end{equation}
  The local characteristic polynomial $P_M(t,u)$ defines a measure density in a vicinity of a point
  $x(t,u)$ in
   tubular neighborhood of oriented hypersurface
  $M$.  If $f(x^a)$ is an arbitrary function which decreases rapidly enough and
  vanishes outside  sufficiently large tubular neighborhood of the surface $M$,  then
    \begin{equation}\label{summarisingformula1}
                \int f(x)d^{n+1}x=\int dt\left(\int f\left(x(u,t)\right)P_M(t,u^i)d\sigma_u\right)\,.
             \end{equation}
 In particular, the polynomial  $P_M(h)$  measures the volume of the half tube $M_h$ if $M$
 is closed  hypersurface.

\section {Dual approach}

  In the previous section  we considered surfaces specified by parametric equations  $x^i=x^i(u)$.
  It is very useful to develop a dual approach,
  i.e., to write the integrals  for hypersurface  defined by an equation $\Phi(x)=0$.
  Formulae written in this language becomes much more transparent  and easier to generalise to supercase.

If we consider a reparameterisation invariant integral $\int
A\left(x,{\p x\over \p u},{\p^2 x\over \p u\p u}\dots\right)d^nu$
over a surface specified by parametric equations $x^i=x^i(u)$,
 then the integrand $A$ is  a {\it density}, which obeys the condition
 $A\left(x,{\p x\over \p \tilde u},{\p^2 x\over \p \tilde u\p \tilde u}\dots\right)=
 A\left(x,{\p x\over \p u},{\p^2 x\over \p u\p u}\dots\right)\cdot\det \left({\p u\over \p\tilde u}\right)$
 if we consider new parameterisation $x(\tilde u)=x(u(\tilde u))$.
 In the dual approach we  come
 to the integral $\int A\left({\p\Phi(x)\over \p x},{\p^2\Phi(x)\over \p x\p x},\dots\right)\delta(\Phi)d^{n+1}x$,
 if a surface is defined by an equation $\Phi(x)=0$.
 The function $A$ in this integral is a {\it dual density}. It obeys the condition:
   \begin{equation}\label{dualdensitycondition}
 A\left({\p\tilde \Phi(x)\over \p x},{\p^2\tilde \Phi(x)\over \p x\p x},\dots\right)\Big\vert_{\Phi=0}=
 G(x)A\left({\p\Phi(x)\over \p x},{\p^2\Phi(x)\over \p x\p x},\dots\right)\Big\vert_{\Phi=0}\,,
\end{equation}
if $\tilde \Phi(x)=G(x)\Phi(x)$. This condition guarantees that
the integral $\int A\delta(\Phi)d^{n+1}x$ does not change if we replace a
function $\Phi$ defining the surface $M$ by the new function $\tilde
\Phi=G\Phi$ ($G\vert_M\not=0$).

\begin{proposition}
  The function
                 \begin{equation}\label{dualdensityforvolume}
   A_{_{\rm vol}}(\p\Phi)=\sqrt {\p_a\Phi\p_a\Phi}
\end{equation}
 defines the dual density corresponding to the volume element at $M$.
 If  the hypersurface $M$ is given by an equation $\Phi(x)=0$, then
 $\vol M=\int A_{_{\rm vol}}(\p\Phi)\,\delta(\Phi)d^{n+1}x$.
                 \begin{equation}\label{dualdensityformeancurvature}
      {\hbox {\it The function}}\qquad A_{_{\rm mcurv}}(\p\Phi,\p^2\Phi)=
    -\p_a\p_a\Phi+{\p_a\Phi\p_b\Phi\p_a\p_b\Phi\over \p_c\Phi\p_c\Phi}
\end{equation}
 defines density corresponding to mean curvature.
 At any point $x$ of the surface $M$ defined by the equation $\Phi(x)=0$
  the ratio $A_{_{\rm mcurv}}/ A_{_{\rm vol}}$ of these densities  is equal
  to the mean curvature  $H(x)$ :
 \begin{equation}\label{menacurvatureindualterms}
    H(\p\Phi,\p^2\Phi)\Big\vert_{x\colon \Phi(x)=0}={A_{_{\rm mcurv}}\left(\p\Phi,\p^2\Phi\right)\over
    A_{_{\rm vol}}\left(\p\Phi\right)}\Big\vert_{x\colon \Phi(x)=0}\,.
\end{equation}
 We have
\begin{equation*}\label{integralsarethesame}
\int A_{_{\rm mcurv}}(\p\Phi,\p^2\Phi)\delta(\Phi)d^{n+1}x=
   \underbrace
   { \int H d\sigma_u=\int n^ag^{kr}x^a_{kr}\sqrt {\det \left(g_{ij}\right)}d^n u}_
   {\hbox {integral of mean curvature over $M$}}\,.
\end{equation*}

\end{proposition}

\begin{proof}
If we replace $\Phi\to G(x)\Phi(x)$, then
$\p_a\Phi\vert_{\Phi=0}\to G\p_a\Phi\vert_{\Phi=0}$ and
$\p_a\p_b\Phi\vert_{\Phi=0}\to G\p_a\p_b\Phi\vert_{\Phi=0}+\p_a
G\p_b\Phi\vert_{\Phi=0}+\p_b G\p_a\Phi\vert_{\Phi=0}$. This implies
that $A_{_{\rm mcurv}}$ and $A_{_{\rm vol}}$ obey condition
\eqref{dualdensitycondition} and are dual densities. To prove that
the density $A_{_{\rm vol}}$ defines volume, note that if
the hypersurface is given by equation $\Phi(x)=x^{n+1}-f(x^1,\dots,x^n)$,
then $\int A_{_{\rm vol}}(\Phi)\delta(\Phi)d^{n+1}x=\int
\sqrt{1+f_1^2+\dots+f_n^2}dx^1dx^2\dots dx^n$.

 Now consider the ratio
 $A_{_{\rm mcurv}}/ A_{_{\rm vol}}$.
 If $\Phi\to G\Phi$, then
 $A_{_{\rm mcurv}}/ A_{_{\rm vol}}\vert_{\Phi=0}$ remains unchanged. Hence
 it is a well-defined function on the surface $M$. For any point on $M$ one can consider
 adjusted Cartesian coordinates in the ambient Euclidean space such that
 $\Phi(x)=x^{n+1}-A_{ij}x^i x^j+o(x^2)$ ($i,j=1,\dots,n$)
  in the vicinity of this point.
 Mean curvature at this point is equal to $H=A_{ii}$. The RHS of
 formula
 \eqref{menacurvatureindualterms} gives the  same answer.
\end{proof}

{\bf Remark} Note that according to general philosophy
   one can come to the dual density corresponding to mean curvature
   taking the variational derivative of the volume functional:
   \begin{equation*}
{\delta\over\delta\Phi}\left(\int  \sqrt{\p_a\Phi\p_a\Phi}\delta
(\Phi)d^{n+1}x\right)=
 -{\p_a\Phi\p_a\Phi\over \sqrt{\p_c\Phi\p_c\Phi}}+{\p_a\Phi\p_b\Phi\p_b\p_a\Phi\over (\p_c\Phi\p_c\Phi)^{3/2}}=
 {A_{_{\rm mcurv}}(\p\Phi,\p^2\Phi)\over A_{_{\rm vol}}(\p\Phi) }
\end{equation*}

 \bigskip

Now we shall find an expression for characteristic polynomial \eqref{charpollocal} in the dual
approach.

Consider the following expression:
\begin{equation}\label{dualdensityforshapeoperator}
    M_{ab}(\p\Phi,\p^2\Phi)=-\p_a\p_b\Phi-{\p_a\Phi\p_b\Phi\p_d\p_d\Phi\over
\p_c\Phi\p_c\Phi}
   +{{\p_a\Phi\p_d\Phi\p_d\p_b\Phi+\p_b\Phi\p_d\Phi\p_d\p_a\Phi
   \over \p_c\Phi\p_c\Phi}}
    \end{equation}
Recall that we do not distinguish between upper and lower indices and implicitly understand summation over
repeated indices.
\begin{proposition}
 Formula \eqref{dualdensityforshapeoperator} defines a matrix-valued dual density.
The ratio of this dual density
   and the dual density $A_{_{\rm vol}}(\p\Phi)=\sqrt {\p_a\Phi\p_a\Phi}$ defines a linear operator
   $\S$ on $\mathbb E^{n+1}$ depending on a point of the surface $M$ defined by an equation $\Phi(x)=0$:
   \begin{equation}\label{shapeoperatorasrationofdensitites}
    {\S}_{ab}(\p\Phi,\p^2\Phi)\Big\vert_{x\colon \Phi(x)=0}=
    {M_{ab}(\p\Phi,\p^2\Phi)\over \sqrt {\p_c\Phi\p_c\Phi}}\Big\vert_{x\colon \Phi(x)=0}\,.
\end{equation}
 The linear operator $\S$ is the direct sum of the Weingarten (shape) operator $S$ acting on vectors tangent to  $M$
and scalar operator of the multiplication by the mean curvature  on vectors
orthogonal to $M$:
             \begin{equation}\label{sumofoperators}
    \S=S\oplus H,\,\, {\rm if}\,\, \v=\v_{_{\rm tangent}}+\v_{_{\rm orthogonal}},\,\, \S\v=
    S\v_{_{\rm tangent}}+H\v_{_{\rm orthogonal}}\,,
                 \end{equation}
where $H=\Tr S$ is mean curvature at point $x\in M$.

  The following relation holds:
\begin{equation}\label{relationbetweendualcharpolynomials}
    \det(1+t\S(x))=\det (1+t S(x))(1+tH(x))
\end{equation}
for an arbitrary point of surface $M$, and the local characteristic
polynomial of the surface $M$ is given by the relation
\begin{equation}\label{charisration}
    P_M(x,t)={\det(1+t\S(x))\over (1+tH(x))}\,.
\end{equation}
\end{proposition}

\begin{proof}
In the same way as above,
one can see that formula \eqref{dualdensityforshapeoperator} defines
a matrix-valued dual density. Hence $\S_{ab}$ is well-defined at the
points $x\colon \Phi(x)=0$ as the ratio of two densities. It is easy
to see that for an arbitrary point $x$ on $M$  in adjusted
Cartesian coordinates $\S_{ij}=A_{ij}$,
$\S_{0 i}=\S_{i 0}=0$ and $\S_{n+1,n+1}=A_{ii}=H$ ($i,j=1,\dots,n$). This implies
\eqref{sumofoperators} and
\eqref{relationbetweendualcharpolynomials}.
\end{proof}

The dual analog of the formula
\eqref{summarisingformula1}
  has the following appearance: $ \int f(x)d^{n+1}x =$
\begin{equation*}
           \int dt
            \left(
      \int f\left(x^a-tn^a(\p\Phi)\right)
             {
            \det
            \left
        (1+t\S(\p\Phi,\p^2\Phi)
          \right)
          \over
          A_{_{vol}}(\p\Phi)+tA_{_{\rm mcurv}}(\p\Phi,\p^2\Phi)
              }A^2_{_{vol}}(\p\Phi)
        \delta(\Phi)d^{n+1}x
           \right),
\end{equation*}
where
        $
     n^a(\p\Phi(x))={\p_a\Phi(x)\over A_{_{\rm vol}} }={\p_a\Phi(x)\over\sqrt {\p_b\Phi(x)\p_b\Phi(x)} }
        $
is unit normal vector field to the surface $\Phi=0$ at the point
$x\colon \Phi(x)=0$.
 (Note that surface defined by the equation $\Phi=0$ is orientable.)

\section {Tube formula for hypersurfaces in superspace}

  Now we analyze how our constructions look in a superspace.
  We will see that the local characteristic function of surfaces in a superspace appeared in the
   tube formula is no longer
   a polynomial. {\it It is a rational function}.

    Consider an $(n+1\vert 2m)$-dimensional Euclidean superspace  with coordinates
   $z^A=(x^a,\theta^\a)$ ($a=1,\dots,n+1$, $\a=1,\dots,2m$),
   where $x^a$ are even coordinates and $\theta^\a$ are odd ones
   ($x^ax^b=x^bx^a$, $x^a\theta^\beta=\theta^\beta x^a$,
   but $\theta^\a\theta^\beta=-\theta^\beta\theta^\a$), with  Riemannian metric $G_{AB}$ such that
      $G_{AB}z^Az^B=x^ax^a+2\theta^1\theta^2+\dots+2\theta^{2m-1}\theta^{2m}$.

A hypersurface, i.e., $(1|0)$-codimensional (or $(n-1|2m)$-dimensional)
surface can be specified
 by parametric equations:
$z^A=z^A(w)$, where $w^I=(u^i,\eta^\mu)=$ $(u^1,\dots,u^{n-1}$;
$\eta^1,\dots\eta^{2m} )$, $u^i$ are even and $\eta^\mu$ are odd
parameters. In the dual approach  a hypersurface can be defined by
an equation $\Phi(z)=0$, where $\Phi$ is an even function.

   Two words about integration in superspace:
   $\int \theta d\theta=1$ and $\int \theta^\a d\theta^\beta=0$ if $\a \not= \beta$.
    Let $f(z)=f(x,\theta)=f_0(x)+f_\a(x)\theta^\a+\dots+f_{_{1\dots q}}(x)\theta^1\dots\theta^{q}$
    be a function on $p|q$-dimensional superspace. Then
    \begin{equation*}
   \int f(z)d^{p+q}z= \int f(x,\theta)d^px d^q\theta=
    \int f_{_{1\dots q}}(x)d^px\,.
\end{equation*}
 The Jacobian of coordinate transformation  $z^A=z^A(\tilde z)$, ($z^A=(x^a,\theta^\a)$,
 is given by Berezinian (superdeterminant) of matrix
 $ \left({\p z^A\over \p \tilde z ^{A'}}\right)=
    \begin{pmatrix}{\p x^a(\tilde x,\tilde \theta)\over \p\tilde x^{a'}} &
          {\p x^a(\tilde x,\tilde \theta)\over \p\tilde \theta^{\a'}}\\
{\p \theta^\a(\tilde x,\tilde \theta)\over \p\tilde x^{a'}} &
          {\p \theta^\a(\tilde x,\tilde \theta)\over \p\tilde \theta^{\a'}}
          \end {pmatrix}$.
(We suppose that all functions of $x$ are smooth and rapidly decreasing at
infinity).

   The Berezinian of an even $p|q\times p|q$ matrix $M$ is given by the following formula
               \begin{equation}\label{berofevenmatrix}
   \Ber M =\Ber
                 \begin{pmatrix}
                        M_{00} &M_{01}\\
                        M_{10}&M_{11}
                 \end {pmatrix}=
                 {\det \left(M_{00}-M_{01}M_{11}^{-1}M_{10}\right)\over \det M_{11}}\,.
\end{equation}
 (Here $M_{00}$, $M_{11}$ are $p\times p$ and $q\times q$ matrices with even entries and
 $M_{01}$, $M_{10}$ are $p\times q$ and $q\times p$ matrices with odd entries.)

   The formulae of previous sections for the first quadratic form, mean
curvature and Weingarten  operator can be easily extended to supercase.
 We just have to be cautious  with sign rule and
consider $\Ber$ instead $\det$.  For example if hypersurface is given by
parameterisation $z^A=z^A(w)$, then
 the first quadratic form is defined by the matrix:
 $g_{_{IJ}}={\p z^A\over\p w^I}G_{AB}{\p z^{B}\over\p w^{J}}(-1)^{p(B)(p(J)+1)}$.
 (By  $p(A)$ we denote the  parity of  corresponding coordinate  $z^A$.)
 The volume element is given by
  $\sqrt  {\Ber g_{_{IJ}}}$ and volume is given by the integral
  $\int \sqrt {\Ber g_{_{IJ}}}d^{2p+q} w$.
  For hypersurfaces the dual density corresponding to the volume form is equal to
  $A_{_{\vol}}=$ $\sqrt {\p_A \Phi G^{AB}\p_B\Phi(-1)^{p(B)}}$.
   Calculations in dual approach for hypersurfaces are typically eaiser.

The tube formula for hypersurface contains a local characteristic function
   \begin{equation}\label{loccharfunction}
   R_M(t,w)=\Ber (1+tS(w))\,.
\end{equation}
The essential difference with previous case
(see \eqref{charpollocal}) is that this local function is no longer a polynomial
in $t$, because  Berezinian  is a rational function of matrix entries.

Recall the following important properties of Berezinian of a linear
operator (see  \cite{vorkhudber}). Let  $A$ be a linear operator in
$p|q$-dimensional space. Let $R_A(t)=\Ber (1+tA)$ be its
characteristic function. (We suppose that $A$ is an even operator.)
 Then
 \begin{itemize}

\item  $R_A(t)=\Ber (1+At)$ is a rational function, the ratio of polynomials of degrees $p$ and $q$ respectively:
\begin{equation}\label{rationalfunctionforber}
R_A(t)=\Ber (1+At)={1+a_1t+\dots+a_pt^p\over 1+b_1t+\dots+b_qt^q}\,.
\end{equation}

\item The expansion of the characteristic function at zero leads to traces of the exterior
powers of the operator $A$:
\begin{equation*}\label{}
    \Ber (1+tA)=\sum^{\infty}_{k=0} c_{k}(A)t^{k} \quad
    \text{where $c_{k}(A)= \Tr \wedge^{k} A$}.
\end{equation*}

\item The expansion of the characteristic function at infinity leads to traces of the
exterior powers of the inverse matrix:
\begin{equation*}
    \Ber (1+tA)=\sum^{\infty}_{k=q-p} c_{-k}^*(A)t^{-k} \quad
    \text{where $c^*_{-k}(A)= \Ber A \cdot\Tr \wedge^{p-q+k} A^{-1}$}.
\end{equation*}

\item The sequences ${c_k}$ ($k=0,1,2,\dots$) and ${c^*_k}$ ($k=p-q,p-q-1,\dots$) are recurrent
sequences with period
$q$.  Moreover the sequence $\gamma_k=c_k-c^*_k$ $k\in \mathbb Z$ is
a recurrent sequence with period $q$.

\item   The following important formula holds:  $\Ber A={\Ber^+ A\over \Ber^-A}$,
where $\Ber^{\pm}$ are invariant polynomial functions  of the
matrix entries of operator  $A$ (in fact polynomials of
$c_k=\Tr\wedge^k A$).
 If  $A$ is presented by a diagonal matrix
 ${\rm diag}[\lambda_1,\dots,\lambda_p;\mu_1,\dots,\mu_q]$, then
                     $$
 \Ber^+ A=R\cdot \prod_{a=1}^p\lambda_a \,, \Ber^- A=R\cdot\prod_{\a=1}^q\mu_a\,,
                      $$
where
               $$
R=\prod_{a=1,\a=1}^{p,q}(\lambda_a-\mu_\a)
               $$
 is resultant of numerator and denominator of the fraction $R_K(t)$.
\end{itemize}
 Note that polynomials arising from direct application of the original formula
 \eqref{berofevenmatrix}  {\it are not}
  invariant polynomials of matrix entries and they have in general degrees $p+pq$ and $q+pq$ respectively.

  Applications of these results to RHS of the tube formula \eqref{loccharfunction}
  gives information about the structure of dufferential-geometrical
  invariants of hypersurfaces in superspace.

Unlike the ordinary case where integration of polynomial function
\eqref{charpollocal} over the surface leads to a polynomial function
\eqref{charpol}, integration of the rational local characteristic function
of a surface in a superspace leads in general to a non-rational function.

\section {Discussion}
Berezinians and  characteristic functions of linear operators in superspace  can
 naturally appear in situations
which originally are not related to anything ''super''.

  Consider the following example.
   Let $A$ be a linear operator in an ordinary linear space $V$.
  Suppose that a linear subspace $M$ of $V$
  is invariant with respect to the action of the operator $A$: $A{\bf v}\in M$ for ${\bf v}\in M$.
   Thus the action of the operator $A$ is well-defined on the factor-space $N=V/M$.
   The characteristic polynomial of the operator $A$ on the factor-space $N$ is equal to the fraction
               \begin{equation*}
    P_{A\vert_{N}}(t)=\det (1+tA\vert_{N})={\det (1+tA\vert_V)\over \det(1+tA\vert_M)}\,.
               \end{equation*}
 One can naturally define an action of the operator $A$ on the superspace $V\oplus \Pi M$,
 where $\Pi$ is parity reversion functor, by setting
 $A (\Pi {\bf v})=\Pi\left(A{\bf v}\right)$. We see that
           \begin{equation*}
P_{A\vert_{N}}(t)= {\det (1+tA\vert_V)\over \det(1+tA\vert_M)}=\Ber
\left(1+tA\vert_{V\oplus \Pi M}\right)= R_{A\vert_{V\oplus \Pi M}}(t)\,.
           \end{equation*}
The rational characteristic function of a linear operator in superspace
naturally appears if we consider operators on factor-space. We have met
this phenomenon for the Weingarten operator of the hypersurfaces
in dual approach (see \eqref{charisration}).
One can say that the characteristic polynomial of the Weingarten operator $S$ on
hypersurface in Euclidean space $\mathbb E^{n+1}$
 can be obtained as the
characteristic function of the operator $\S$ extended on  $(n+1\vert
1)$-dimensional superspace.

In this example the fraction is reducible.  Numerator and
denominator of the fraction $R_{A\vert_{V\oplus \Pi M}}$ contain a
common factor, the polynomial $P_{A\vert_{M}}$.

\begin{proposition}
 Let $A$ be a linear operator on a superspace $V$ and $M$ be an invariant subspace of the operator $A$.
Then the characteristic functions of the operator $A$ on superspace
$V/M$ and superspace $V\oplus \Pi M$ coincide:

\begin{equation}\label{eulercharcteristic}
R_{A\vert_{V/M}}(t)= R_{A\vert_{V\oplus  \Pi M}}(t)\,.
\end{equation}
\end{proposition}
This simple but important statement demonstrates that
a characteristic function can be considered as a multiplicative
version of Euler characteristic.
 It is this property of Berezinian
which makes it an adequate tool for describing  Reidemeister
torsion.
  Let us recall its construction.
  Consider a complex $(E=E_0\oplus E_1,d)$ as a superspace. Here the differential $d$ is an odd operator.
   Denote by $Z$ the kernel of the operator $d$ and by $B$, its image.
   Then the cohomology of $d$ is $H=Z/B$.  Denote by $\Ber(V)$ the space of volume forms on superspace $V$.
   Then  $\Ber(Z)=\Ber(H)\otimes \Ber (B)$ and $\Ber (E)=\Ber (Z)\otimes \Ber (\Pi B)$,
   because differential $d$ is an odd linear operator. Hence
   the space $\Ber (E)$ is canonically isomorphic to the space $\Ber(H)$.
   Reidemeister torsion can be understood as this canonical isomorphism
   \footnote
    {This construction was studied
   by A.S.\,Schwarz and applied by him in particular to partition function of degenerate quadratic
   functional in Quantum Field Theory (see \cite{schwarz2}).}.

  We can say something more.

  \begin{proposition}
  For an arbitrary (even) operator $A$ on complex
  $E$ which commutes with
  differential $d$  the following equality  holds:\begin{equation*}
    R_{A\vert_{H}}(t)=R_{A\vert_{E}}(t)\,.
\end{equation*}
\end{proposition}
\begin{proof}
According to \eqref{eulercharcteristic}
$R_{A\vert_{H}}(t)=R_{A\vert_{Z/B}}(t)=R_{A\vert_{Z\oplus \Pi B}}(t)=R_{A\vert_{E}}(t)$.
\end{proof}

 Our considerations reveal that a rational function $R(t)$ such that $R(1)=1$ can be interpreted
as the characteristic function of the linear operator in a superspace.
Furthermore if we interpret a linear operator as the Weingarten
operator of a surface in a superspace, then this rational function
can be considered as a density of supervolume of a tube.
\medskip

       \begin{tabular}{|p{ 2cm}|}
           \hline
$A$ is linear operator in superspace\\
        \hline
      \end{tabular}
$\longleftrightarrow$
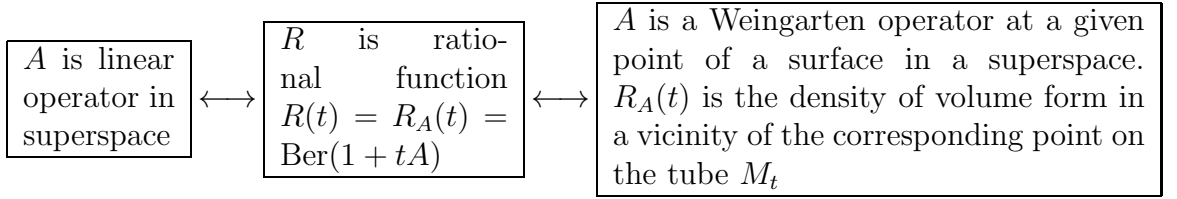
          \begin{tabular}{|p{3cm}|}
            \hline
$R$ is rational function
$R(t)=R_A(t)=\Ber (1+tA)$  \\
   \hline
\end{tabular}
$\longleftrightarrow$
\begin{tabular}{|p{7cm}|}
             \hline
  $A$ is a Weingarten operator at a given point  of a surface in a superspace.
  $R_A(t)$ is the density of volume form in a vicinity of the corresponding point on the tube $M_t$
  \\
  \hline
\end{tabular}

\bigskip

Let us consider an example of a different origin.

Let $X$ be an arithmetic variety given by polynomial $P_X$ in $n$
variables with coefficients in a finite field, say $\mathbb F_p$
($p$ is prime number).
 Denote by $\nu_k$ number of  points of $X$ over the field extension
 $\mathbb F_{p^k}\supset \mathbb F_p$,
 i.e. number of solutions of the equation $P_X=0$
 in the space $\mathbb F_{p^k}^n$. The zeta-function of arithmetic
surface can be defined as
  \begin{equation}\label{zeta-functiondef}
    Z_X(t)=\exp{\sum_{k=0}^\infty {\nu_k\over k}t^k}
\end{equation}
  (see for e.g.  book \cite{maninshafar}).
 One of  the deep results in algebraic number theory is that $Z_X(t)$ is rational function of argument $t$.
 It is the first of the famous Weil conjectures proved by Dwork in 1960.

  In view of the above we can suggest that this rational function is a characteristic function of a
  linear operator in a superspace. The properties expressed in Propositions 3 and 4 reveal a cohomological
  interpretation of this operator.
   Furthermore one can interpret this characteristic function as the volume density of
   a ``tubular neighborhood``, i.e. an analogue of Weyl tube formula.
   Philosophically it should not be a surprise, since the definition \eqref{zeta-functiondef}
   of zeta-function can be seen as a formula for ``volume`` of a formal neighborhood of a single point.
    The whole surface $\bar X\supset X$
   over algebraic closure of $\bar{\mathbb F}_p\supset \mathbb F_p$ can be viewed as a ``tubular neighborhood``
   of this single point.  A full  understanding of this relation is yet to be achieved.

\section*{Acknowledgements}
  I am deeply grateful to Th.Th.\,Voronov for valuable comments and for encouraging for writing this note.
  I am also grateful to A. Haunch for interesting discussions.



\begin{thebibliography}{99}\itemsep=-.2pc


\bibitem{candelas}  Ph.\, Candelas, X. de la Ossa.
{\it The Zeta-Function of a p-Adic Manifold, Dwork Theory for
Physicists},  arXiv:hep-th 0705.2056

\bibitem{grey} A.\,Gray.   {\it Tubes},  Addison--Wesley Publishing Company, USA, 1990

\bibitem{vorkhudber}
H.M.\,Khudaverdian,  Th.Th.\,Voronov.  {\it Berezinians, exterior powers and
recurrent sequences.}
 Lett. Math. Phys., 74(2):201--228, 2005.

\bibitem{schwarz2}    A. Schwarz.
 {\it Semiclassical approximation in Batalin-Vilkovisky formalism},
 Commun.Math.Phys. 158 (1993) 373-396

\bibitem{maninshafar}


I.R.\,Shafarevich. {\it Basic algebraic geometry. 1. Varieties in
projective space. 2nd edition.} Springer-Verlag, Berlin, 1994.

--







\end{thebibliography}
\end{document}